\documentclass[dvips]{article}

\usepackage{icrc2011}

\title{LUNASKA simultaneous neutrino searches with multiple telescopes}

\newcommand{\etal}{\MakeLowercase{\textit{et al. }}} 
\newcommand{\url}[1]{\texttt{#1}}

\shorttitle{J.~D. Bray \etal LUNASKA neutrino search}

\authors{J.~D. Bray$^{1,2}$, R.~D. Ekers$^2$, C.~W. James$^3$, P. Roberts$^2$, A. Brown$^2$, C.~J. Phillips$^2$, R.~J. Protheroe$^1$, J.~E. Reynolds$^2$, R.~A. McFadden$^4$, M. Aartsen$^1$}

\afiliations{$^1$School of Chemistry \& Physics, Univ. of Adelaide, SA, Australia\\ $^2$CSIRO Astronomy \& Space Science, Epping, NSW, Australia\\ $^3$Department of Astrophysics, IMAPP, Radboud University Nijmegen, The Netherlands\\ $^4$ASTRON, 7990 AA Dwingeloo, The Netherlands}

\email{justin.bray@adelaide.edu.au}

\abstract{The most sensitive method for detecting neutrinos at the very highest energies is the lunar Cherenkov technique, which employs the Moon as a target volume, using conventional radio telescopes to monitor it for nanosecond-scale pulses of Cherenkov radiation from particle cascades in its regolith.  Multiple-antenna radio telescopes are difficult to effectively combine into a single detector for this purpose, while single antennas are more susceptible to false events from radio interference, which must be reliably excluded for a credible detection to be made.  We describe our progress in excluding such interference in our observations with the single-antenna Parkes radio telescope, and our most recent experiment (taking place the week before the ICRC) using it in conjunction with the Australia Telescope Compact Array, exploiting the advantages of both types of telescope.}


\begin{document}

\maketitle

\section{Introduction}

The particle cascades induced in dense media by UHE (ultra-high energy; $>$EeV) cosmic rays and neutrinos may be detected through the Askaryan effect~\cite{askaryan1962}, which results in a nanosecond-scale coherent radio pulse.  Dagkesamanskii and Zheleznykh suggested that ground-based radio telescopes could be used to monitor the upper layers of the Moon, which are a suitable radio-transparent target material, to use the Moon as a neutrino detector~\cite{dagkesamanskii1989}.

Compared to similar experiments exploiting the Askaryan effect to search for UHE particles in terrestrial volumes of ice~\cite{gorham2010} or salt~\cite{connolly2010}, this technique has a larger potential aperture (as the Moon is a large target), but has a higher threshold energy for a particle to be detectable (as the Moon is far away).  It also has the advantage that suitable instruments, in the form of radio telescopes, already exist; but these are also used for other experiments, so they are only intermittently available for use for UHE particle detection.

A significant difficulty with this technique is distinguishing between a real event and a spurious pulse of RFI (radio frequency interference) produced by artificial electrical equipment.  Without the ability to do this reliably, it is possible to set a limit to the UHE particle flux based on the strength and/or number of observed pulses, but it is not possible to confidently attribute a particular pulse to a particle cascade, and hence to claim the detection of a UHE particle.

\section{Types of experiment}

Current experiments employing this technique have several factors in common.  They all monitor the signal from a radio telescope in real time, keeping a buffer of the most recent data.  If a potential event is detected, this triggers the permanent storage of the buffered data, which can then be examined in more detail.

These experiments vary, however, in the details of the real-time algorithm used to identify a potential event.  Apart from steps including dedispersion, interpolation, etc., they can be divided into two types according to whether they use their radio telescope, which may consist of one or several antennas, in an incoherent or a coherent fashion.

\subsection{Incoherent experiments}

An incoherent experiment makes use of a radio telescope consisting of multiple antennas, each of which acts as an independent detector.  The radio pulse from a UHE particle interaction on the Moon will be seen by all antennas, with relative arrival times indicating its arrival direction.  Experiments of this type include RESUN~\cite{jaeger2010} and previous LUNASKA observations with the ATCA (Australia Telescope Compact Array)~\cite{james2010}.

The requirement that a pulse must be seen by all antennas, with relative arrival times indicating that it originated from the Moon, provides an effective discriminant against false events from RFI.  However, the sensitivity of this type of experiment is limited by the sensitivity of a single antenna.

\subsection{Coherent experiments}

This type of experiment coherently combines the incoming radio waves across the entire collecting area of a radio telescope before checking the signal for a possible pulse.  The telescope may consist of either a single large antenna or an array.  In the latter case, the signals from separate antennas must be combined electronically, which requires special-purpose digital hardware.

Due to the larger effective collecting area, this approach results in improved radio sensitivity, allowing it to detect pulses from lower-energy UHE particles.  However, a larger antenna or array of antennas forms a narrower beam, allowing it to observe only a fraction of the Moon and reducing its aperture to UHE particles.  This drawback can be offset by using multiple beams directed at different parts of the Moon.  For a single antenna, this is achieved by placing multiple receiver feeds at the antenna focus.  For an array, it involves combining the signals from separate antennas at a different set of relative delays for each beam.

A real event should appear as a pulse in only one beam: the beam directed towards the region of the Moon where it occurred.  This is used to exclude false events from RFI, which generally appear in multiple beams.  However, previous experiments have found this to be a weaker discriminant than the RFI rejection scheme used in incoherent experiments, whether using a single antenna~\cite{bray_inprep} or an array~\cite{buitink2010}, so they have been limited by the RFI background.

\section{This experiment}
\label{sec:our_exp}

The work described here is a development of a previous experiment of the coherent type described above, using a single large antenna: the 64 metre Parkes radio telescope.  The previous experiment used four beams, three being directed at the Moon, with the fourth used purely for RFI rejection.  One beam was always on the limb of the Moon closest to Centaurus~A, to maximise directional sensitivity to UHE particles from that potential source~\cite{abreu2011}, as in a previous experiment~\cite{james2011}.  More pulses were observed than expected from Gaussian noise, which may be due to remnant RFI; but the possibility of real events cannot be excluded~\cite{bray_inprep}.

\begin{figure}
 \begin{center}
  \includegraphics[width=\linewidth]{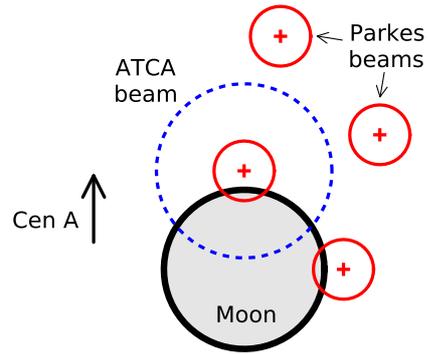}
 \end{center}
 \caption{Typical beam pointings relative to the Moon (shaded) for the Parkes (solid) and ATCA (dashed) telescopes.  Crosses indicate the orientation of the orthogonal linear polarisations of the Parkes receiver.  The two off-Moon Parkes beams are for RFI rejection; the other beam positions are a compromise between exposure to the Moon, and reducing the thermal noise received from the Moon to increase the sensitivity.  The ATCA beam is placed to be able to confirm a possible detection in the Parkes beam sensitive to UHE particles from the direction of Centaurus A.}
 \label{fig:pointing}
\end{figure}

The purpose of this experiment is to determine whether this excess of observed pulses includes any genuine UHE particle interaction events, by observing simultaneously with a second telescope and searching for coincident pulses.  The second telescope was the ATCA, an array of which we used five antennas in a compact configuration.  The frequency range of the ATCA receivers was 1.1--3.1 GHz, which encompasses the 1.2--1.5 GHz range of the Parkes receiver.  The ATCA antennas were pointed at the same location as a single Parkes beam, as shown in figure \ref{fig:pointing}.

To match the sensitivity of the Parkes telescope, we require a coherent combination of the signals from the five ATCA antennas.  However, we were not able to form, in real time, the multiple combined signals that would be required.  Instead, we used the detection of a pulse at Parkes as a trigger to store buffered data on all ATCA antennas, which can then be retroactively combined to search for a coincident pulse.

Essentially, this experiment uses each of the involved telescopes in a coherent manner: the Parkes telescope is a single antenna, so it is inherently coherent, while the multiple antennas of the ATCA are coherently combined.  The combination between the two telescopes, however, is incoherent: the signal from each telescope is examined for a pulse independently.  The sensitivity is similar to the previous experiment using the Parkes telescope alone, as a pulse must exceed the trigger threshold on this telescope in order to be recorded.  However, the use of a second telescope to confirm potential events should allow complete rejection of the RFI background, allowing much greater confidence if a coincident pulse is found.

This experiment was conducted during 4-7 August, immediately prior to the 32nd International Cosmic Ray Conference.  We have not yet determined whether there were any coincident pulses between the Parkes and ATCA telescopes during the observation period.  However, we have established that we were successful in performing several types of calibration required for this type of experiment.

\section{Calibration}

This experiment required careful calibration of the delays between the antennas of the ATCA, and between the Parkes and ATCA telescopes, which are separated by 300 km.  The first of these is required for normal use of the ATCA, and the second is a routine procedure in VLBI (Very Long Baseline Interferometry).  However, in our case they must be accomplished using only the short buffers which are captured in this experiment.

\subsection{ATCA calibration}

To coherently combine the signals from separate antennas of the ATCA, the relative delays must be calibrated to within a small fraction ($\sim 1/20$) of the inverse of the bandwidth.  Since we aim to find pulses corresponding to those detected by the Parkes telescope in the frequency range 1.2--1.5 GHz, we require timing precision of $\sim 0.2$ ns.

Achieving this requires cross-correlation between antennas of a broad-band signal such as noise from an astronomical source.  For this purpose we used the quasar 3C273, as it was unresolved by our array, and bright enough for a strong cross-correlation within the maximum buffer size that could be captured with our system (6 ms).

\begin{figure}
 \begin{center}
  \includegraphics[width=\linewidth]{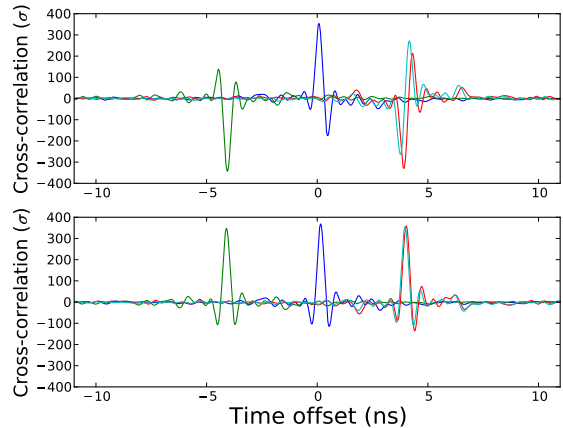}
 \end{center}
 \caption{Sample cross-correlations between antenna 1 of the ATCA and the other four antennas, from a single 6 ms buffer while observing 3C273.  Fourier interpolation has been performed, and only one polarisation is shown.  Without phase corrections (top panel), the exact time offset is unclear.  With phase corrections (bottom panel), the peaks are well-defined, and their positions are stable to $\pm 0.09$ ns between this and other buffers.  The structure away from the peaks, which deviates from an idealised $\mathrm{sinc}$ function, may be due to the extended structure of 3C273, or other sources in the field of view.}
 \label{fig:atcacal}
\end{figure}

These buffers of baseband data also allow calibration of the relative phase of each antenna, which is also required in order to coherently combine the data from multiple antennas.  In fact, this phase calibration is required in order to obtain delays which meet our precision criterion, as shown in figure \ref{fig:atcacal}.  Phase calibration can be checked for consistency via the requirement that the phases around any closed loop of antennas must sum to zero: applying this, we find that these sums vary from zero by $\pm 2.5^\circ$, which indicates negligible error.

\subsection{Parkes-ATCA calibration}

It was necessary to calibrate the relative timing between the Parkes and ATCA telescopes for two reasons.  The first is to be able to determine, if a pulse is found in the ATCA data corresponding to a pulse detected at Parkes, whether the arrival time indicates that it originated from the Moon.  If the timing offset is known precisely, there is a $\sim 4$ $\mu$s window within which the ATCA pulse could arrive, corresponding to the width of the Parkes beam.  If there is uncertainty in the timing offset, this search window must be extended, resulting in a higher threshold for significance and hence reduced sensitivity.

The second reason for this calibration is to ensure that the correct data are stored.  Depending on the position of the Moon on the sky, the relative arrival times of a pulse could vary by about 1 ms.  In normal operation, however, we only store a 200 $\mu$s segment of the buffer at the ATCA.  This segment must be selected from the buffer based on the current position of the Moon.  The precision required for this purpose is only $\pm$ 100 $\mu$s, to ensure that the correct time is within the range of the stored data.

This calibration is performed in a similar way to the ATCA internal calibration, but with a cross-correlation between buffers captured at the Parkes and ATCA telescopes.  Our system at the Parkes telescope has a maximum buffer size of 8 $\mu$s, considerably less than at the ATCA, which reduces the strength of the correlation and requires that we use a stronger calibrator source.  In addition, the source must be compact enough to be unresolved on the 300 km Parkes-ATCA baseline, which limits the range of suitable astronomical objects.

\begin{figure}
 \begin{center}
  \includegraphics[width=\linewidth]{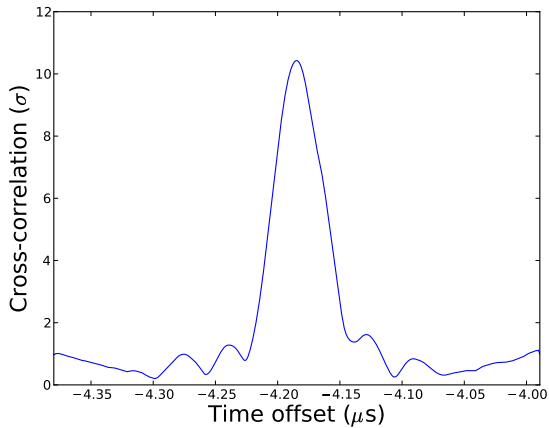}
 \end{center}
 \caption{Cross-correlation between Parkes and antenna 1 of the ATCA from a single pair of corresponding buffers while observing BeiDou-1C, discarding ATCA data from outside the Parkes frequency range.  The result has been rectified and smoothed with a Gaussian of width 10 ns.  The time offset is relative to the arrival time predicted by the satellite ephemeris; the 4.2 $\mu$s offset is consistent between observations and hence is a systematic effect which can be removed.  The width of the peak indicates that we have achieved timing precision of $\sim$ 50 ns.}
 \label{fig:intercal}
\end{figure}

Due to the lower required precision, however, a narrow-band signal is sufficient for this purpose.  As our source, we used BeiDou-1C, a Chinese navigational satellite which emits strongly in the range 1220-1245 MHz.  We determined its position with a TLE-format ephemeris~\cite{sats} and the PyEphem astronomical library~\cite{pyephem}.  The timing precision we achieved, shown in figure \ref{fig:intercal}, is $\sim$ 50 ns, which corresponds roughly to the theoretical limit of the inverse bandwidth.  There will be an additional error from uncertainty in the satellite ephemeris, which we expect to be less than this.

\section{Lunar satellite interference}

\begin{figure}
 \begin{center}
  \includegraphics[width=\linewidth]{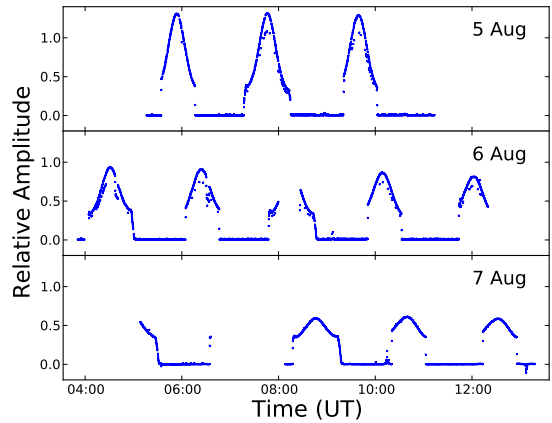}
 \end{center}
 \caption{Intensity of narrow-band interference at 2.27 GHz, relative to the underlying noise spectrum.  The behaviour is as expected for a lunar satellite: it peaks as the satellite orbits through the telescope beam, and falls to zero when the satellite is occulted by the Moon.  The variation from one day to the next is due to the changed pointing of the telescope (see section \ref{sec:our_exp}) relative to the fixed orbit of the satellite.}
 \label{fig:time}
\end{figure}

We detected narrow-band interference at a frequency of 2.27 GHz, varying with a period of 133 minutes (see figure \ref{fig:time}).  These figures correspond respectively to the communications frequency and orbital period of the NASA Lunar Reconnaissance Orbiter (LRO)~\cite{lro}.

This narrow-band interference is simple to remove from the data.  However, there is the possibility that an electrical system on the LRO (or another lunar satellite) may be capable of producing nanosecond-scale broad-band radio pulses.  Unlike terrestrial interference, such a pulse would be seen by both telescopes used in this experiment, with relative arrival times indicating that it originated from the Moon.  This makes it difficult to distinguish from a real UHE particle-induced Askaryan pulse.

We are investigating the feasibility of such pulses being generated by the hardware of a lunar satellite.  If this is feasible, then it will be necessary to compare the position of any detected pulse, as determined from its arrival times, with the simultaneous position of all lunar satellites.  The position of a satellite may be obtained either from a published ephemeris, or from interferometry on its communications signal.

\section{Conclusion}

We have demonstrated the ability to search for lunar Askaryan pulses with two radio telescopes simultaneously, with good timing calibration between them.  This allows coincidence detection to confirm the lunar origin of a detected pulse, which is a prerequisite for our experiment to reliably detect a UHE particle interaction on the Moon.

\section{Acknowledgements}

The Australia Telescope Compact Array and Parkes telescope are both part of the Australia Telescope which is funded by the Commonwealth of Australia for operation as a National Facility managed by CSIRO.  This research was supported by the Australian Research Council’s Discovery Project funding scheme (project number DP0881006).

\clearpage

\end{document}